\documentclass[aps,prl,twocolumn,superscriptaddress]{revtex4-2} 
\usepackage{mathrsfs}
\usepackage{amsfonts}
\usepackage{amsmath}
\usepackage{txfonts}
\usepackage{amssymb}
\usepackage{graphicx,subfigure}
\usepackage{bm}
\usepackage{color}
\usepackage[normalem]{ulem}
\usepackage{dcolumn} 
\usepackage{bbold}
\usepackage[table]{xcolor}
\usepackage{float}
\usepackage{tabularx}
\newcommand{\ket}[1]{|#1\rangle}
\newcommand{\bra}[1]{\langle #1|}
\newcommand{\Tr}{\text{Tr}}

\newcommand{\B}[1]{\textcolor{blue}{#1}}

\begin{document}

\title{Graph-Spectral Entanglement} 

\author{Vahid Azimi-Mousolou}
\email{Electronic address: vahid.azimi-mousolou@physics.uu.se}
\affiliation{Department of Physics and Astronomy, Uppsala University, Box 516, 
SE-751 20 Uppsala, Sweden}

\date{\today}

\begin{abstract}
We introduce a unified graph-spectral framework for quantifying multipartite entanglement, built upon an entanglement graph and its associated entanglement matrix. The framework accommodates general subsystem and partition structures of composite quantum systems, while offering a complementary geometric and network perspective on characterizing multipartite quantum correlations. We prove that the resulting graph-spectral entanglement satisfies the requirements of an entanglement measure. Furthermore, we derive a multipartite monogamy relation and, based on the resulting spectral hierarchy, introduce residual graph-spectral entanglement for arbitrary multipartite states. This establishes a general construction of residual entanglement beyond conventional qubit-based formulations and provides a new approach to investigating, quantifying, witnessing, and classifying multipartite entanglement in both theoretical and experimental settings.  
\end{abstract}

\maketitle

Quantum entanglement, originally introduced in foundational debates \cite{Einstein1935, Schrodinger1935, Schrodinger1936}, is now recognized as a central resource in quantum information science, enabling advantages in quantum communication, computation, and metrology \cite{Horodecki2009, Nielsen2010, Giovannetti2011, Chitambar2019}. 
In particular, multipartite entanglement plays an essential role in measurement-based quantum computation \cite{Raussendorf2001, Raussendorf2003}, quantum networks \cite{Cirac1997, Briegel1998, Kimble2008, Simon2017, Epping2017, Wehner2018, Kraft2021, Navascues2020}, and multiparty quantum cryptography, including quantum key distribution and secret sharing \cite{Gaertner2007, Zhou2018, Proietti2021, Yang2024}.
While bipartite entanglement has been extensively studied within a relatively mature theoretical framework, the characterization of multipartite entanglement remains challenging \cite{Navascues2020, Ma2024}. The difficulty arises not only from the rapidly increasing dimension of the Hilbert space, but also from the rich hierarchy of inequivalent correlation structures that emerge when more than two subsystems are involved \cite{Dur2000,Verstraete2002}.

Various multipartite entanglement measures have been developed to characterize different aspects of this structure. Residual entanglement captures multipartite contributions through monogamy relations among bipartite entanglement contributions \cite{Coffman2000,Osborne2006, Lohmayer2006}. Genuine multipartite concurrence characterizes multipartite entanglement through a minimization over bipartitions \cite{MaGME2011}, while the Schmidt measure quantifies multipartite entanglement through the minimum number of product terms in a state decomposition, generalizing the Schmidt rank \cite{Eisert2001}. Geometric measures quantify the distance or overlap of a state with the closest unentangled state \cite{Wei2003}, while global entanglement measures characterize multipartite entanglement through scalar functions of reduced-state purities \cite{Meyer2001}.
Multipartite entanglement has also been investigated through entanglement witnesses, semidefinite-programming methods, convex-roof constructions, and resource-theoretic approaches \cite{Guhne2009,Jungnitsch2011,Chitambar2019,Ma2024}. These approaches are complementary, emphasizing different aspects of multipartite entanglement.

Bipartite entanglement is well characterized by a set of measures, including concurrence, entanglement of formation, relative entropy of entanglement, negativity, and squashed entanglement \cite{Hill1997,Wootters1998,Vedral1998,Vedral1998, Rungta2001, Vidal2002, Christandl2004, Plenio2007}, subject to general consistency requirements \cite{Vedral1997,Vidal2000, Donald2002}. However, multipartite entanglement involves correlations among multiple subsystems and exhibits a richer structure that cannot, in general, be reduced to a single bipartite quantity. Even for few-body systems, inequivalent entanglement classes arise \cite{Dur2000,Verstraete2002}, and their classification becomes increasingly challenging with system size \cite{Gour2013}. Existing multipartite entanglement measures emphasize different aspects of this structure through different ways of combining information associated with the many-body system and its subsystem-partitioning structure. This diversity reflects the complexity of multipartite entanglement and motivates frameworks that can represent such information within a common description. In particular, a perspective that retains the correlation structure encoded in both the strength and connectivity of bipartite correlations provides a distinct way to characterize multipartite entanglement.

The broader relevance of multipartite entanglement is reflected in quantum resource theories, quantum metrology, and many-body quantum systems \cite{Chitambar2019,Toth2012,Friis2018}. Multipartite entanglement is relevant to enhanced phase-estimation sensitivity \cite{Toth2012} and becomes particularly important in many-body systems with multiple experimentally accessible subsystems, where a characterization based on a fixed partition may not fully reflect the underlying correlation structure. 
For example, multipartite entanglement has been experimentally characterized in a fully controlled 20-qubit system, in which different groups of qubits develop multipartite correlations through many-body dynamics \cite{Friis2018}, and in a system of 25 individually accessible atomic quantum interfaces, in which genuine 9-, 14-, and 22-partite entanglement was demonstrated \cite{Pu2018}. Furthermore, large-scale multipartite entanglement has also been experimentally demonstrated between more than 200 atomic ensembles and among millions of atoms in a solid \cite{Zarkeshian2017,Frowis2017}.
Such settings motivate frameworks that can accommodate different subsystem groupings while retaining information about their collective quantum correlation structure.

In this Letter, we introduce a unified graph-spectral framework for multipartite entanglement quantification based on an entanglement graph and its associated entanglement matrix. The framework accommodates different subsystem and partition structures and allows different bipartite entanglement measures to define the edge weights within a common graph-spectral construction. The spectral radius of the entanglement graph provides an aggregation principle that captures the dominant collective correlation structure of the entanglement-weighted network, offering a perspective distinct from existing approaches. We prove that the resulting spectral entanglement measure satisfies the standard requirements of an entanglement measure. We further establish an inclusion monotonicity for the spectral construction, which yields a generic spectral analog of a monogamy inequality that is structurally distinct from conventional algebraic monogamy relations \cite{Coffman2000,Osborne2006}. Based on the resulting spectral hierarchy, we introduce a residual graph-spectral entanglement for arbitrary multipartite states. The framework is also relevant to experimentally controlled many-body systems, where entanglement can be characterized across different groups of subsystems and partition structures \cite{Friis2018,Lu2018}. Beyond quantification, the graph-spectral approach provides a complementary geometric and network perspective on multipartite quantum correlations.

\begin{figure}[H]
    \centering \includegraphics[width=0.32\textwidth]{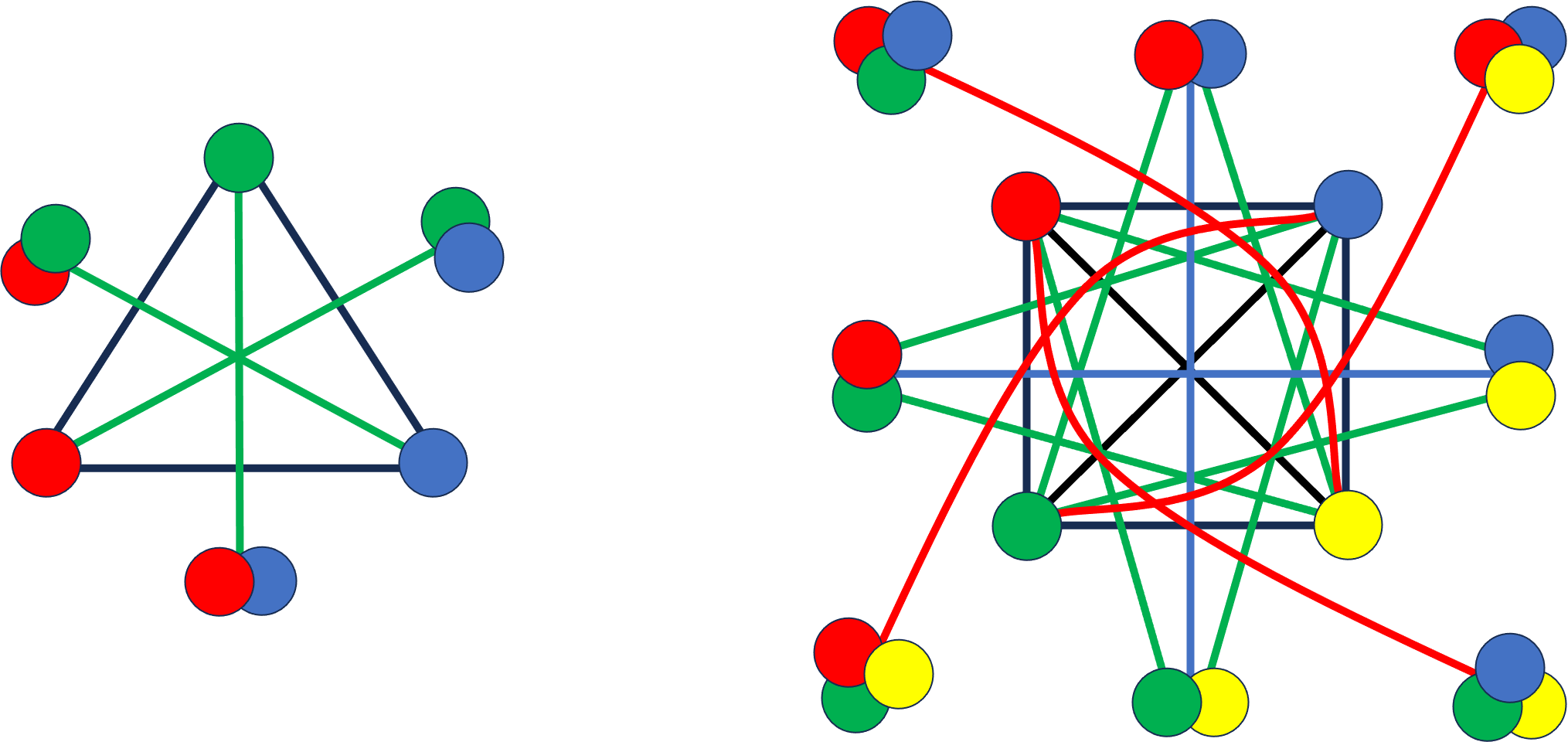}
    \caption{Schematic illustrations of multipartite entanglement graphs for three- and four-body systems. Colored discs represent constituent subsystems, with overlapping discs indicating subsystems grouped into a partition block. Edges connect disjoint partition blocks and represent the bipartite entanglement between them. The edge color identifies the type of bipartition rather than the magnitude of the corresponding edge weight. The three-body graph illustrates the $1$-$1$ (black) and $1$-$2$ (green) bipartitions, while the four-body graph illustrates the $1$-$1$ (black), $1$-$2$ (green), $1$-$3$ (red), and $2$-$2$ (blue) bipartitions, highlighting the increasing complexity of the multipartite correlation structure with system size.green
    }
    \label{fig:GR1}
\end{figure}

The key idea is to represent the multipartite entanglement structure of a quantum state as a weighted graph whose nodes correspond to subsystems or partitions and whose edges encode entanglement between disjoint nodes.
As illustrated in Fig.~\ref{fig:GR1} for three- and four-body systems, this representation naturally captures the full connectivity of multipartite quantum correlations. The spectral characteristics of the resulting graph provide a unified and physically meaningful quantification of entanglement in many-body quantum systems.

\textit{Definition.---}
Consider an arbitrary mixed state $\rho$ of an $N$-partite quantum system whose
total Hilbert space is 
\begin{equation}
\mathcal{H}=\bigotimes_{i\in I}\mathcal{H}_i,
\qquad I=\{1,2,\ldots,N\},
\label{eq:NPS}
\end{equation}
where each $\mathcal{H}_i$ denotes the Hilbert space of subsystem $i\in I$.
Let $1<n \le 2^{N}-2$ and $1<M\le N$, and consider a family of nontrivial
subsets of $I$,
\begin{equation}
\Lambda=\{I_1,I_2,\ldots,I_{n}\}
\subseteq \mathcal{P}(I)\setminus\{\emptyset,I\},
\label{MDSS}
\end{equation}
where $\mathcal{P}(I)$ denotes the power set of $I$ and
\begin{equation}
I_{\Lambda}:=\bigcup_{k=1}^{n} I_k=\{i_1,\ldots,i_M\}\subseteq I .
\label{MDSS1}
\end{equation}

For a given bipartite entanglement measure $E$~\cite{Horodecki2009, Hill1997,Wootters1998,Vedral1998,Vedral1998, Rungta2001, Vidal2002, Christandl2004, Plenio2007}, we define the \emph{entanglement graph} $G(\rho,\Lambda,E)$ associated with the state $\rho$ and subsystem family $\Lambda$ as follows. As shown in Fig.~\ref{fig:GR1} for two illustrative cases, each vertex associated with an element $I_k\in\Lambda$ represents a composite subsystem whose Hilbert space is
\begin{equation}
\tilde{\mathcal{H}}_{I_k}
=\bigotimes_{i\in I_k}\mathcal{H}_i.
\end{equation}
The edge weight between two vertices $I_k$ and $I_l$ is then given by the bipartite entanglement between the corresponding subsystems, i.e.,
\begin{equation}
E_{I_k|I_l}=
\begin{cases}
E(\rho_{I_k I_l}), & \text{if } I_k\cap I_l=\varnothing,\\
0, & \text{otherwise}.
\end{cases}
\end{equation}
Here
$\rho_{I_k I_l}
=\mathrm{Tr}_{I\setminus(I_k\cup I_l)}[\rho]$
denotes the reduced density matrix on $\tilde{\mathcal{H}}_{I_k\cup I_l}$ obtained by tracing out all subsystems $\mathcal{H}_i$ with $i\notin I_k\cup I_l$.

The weighted adjacency matrix of the entanglement graph $G(\rho,\Lambda,E)$, defined as $\mathcal{M}_{E,\Lambda}(\rho)=[m_{kl}]_{k,l=1}^{n}$, where $m_{kl}=E_{I_k|I_l}$, is referred to as the \emph{entanglement matrix}.
In the following, we show that $\mathcal{M}_{E,\Lambda}(\rho)$ provides a natural way to quantify $M$-partite entanglement among the subsystems $\mathcal{H}{i_1}, \ldots, \mathcal{H}{i_M}$ in the state $\rho$ with respect to the specified family $\Lambda$.

The entanglement matrix $\mathcal{M}_{E,\Lambda}(\rho)=[m_{kl}]_{k, l=1}^{n}$ is real, entrywise nonnegative, symmetric, and hollow~\cite{Horn2012}, i.e.,
\begin{equation}
m_{kl}=E_{I_k|I_l} \ge 0, \quad
\mathcal{M}_{E,\Lambda}(\rho) = \mathcal{M}_{E,\Lambda}^{\mathsf{T}}(\rho), \quad
\Tr[\mathcal{M}_{E,\Lambda}(\rho)] = 0,
\label{MEM}
\end{equation}
which implies that all eigenvalues $\{\lambda_k\}_{k=1}^{n}$ are real and satisfy $\sum_{k=1}^{n} \lambda_k = 0$. This allows us to define the $M$-partite  \emph{graph-spectral entanglement} (GSE) among the subsystems $\mathcal{H}_{i_1},\ldots,\mathcal{H}_{i_M}$ as the largest eigenvalue of the entanglement matrix \cite{note}
\begin{equation}
E_\Lambda(\rho) = \lambda_{\max}\bigl[\mathcal{M}_{E,\Lambda}(\rho)\bigr].
\label{eq:multipartiteE}
\end{equation}
The associated separability structure is defined in the End Matter below.

Using the properties of the entanglement matrix in Eq.~\eqref{MEM} and the Rayleigh quotient~\cite{Horn2012}, the GSE can be equivalently expressed as

\begin{itemize} 
    \item[] \textit{Symmetric quadratic form:}
    \begin{equation}
    E_\Lambda(\rho) = \max_{\|{\bf x}\|_2=1} {\bf x}^\dagger \mathcal{M}_{E,\Lambda}(\rho) {\bf x}, 
    \qquad {\bf x}\in \mathbb{R}^{n},
    \label{eq:quadraticformmultipartiteE}
    \end{equation}

    \item[] \textit{Spectral radius:}
    \begin{equation}
    E_\Lambda(\rho) = r\bigl[\mathcal{M}_{E,\Lambda}(\rho)\bigr] := \max_i \bigl|\lambda_i[\mathcal{M}_{E,\Lambda}(\rho)]\bigr|,
    \label{SRG}
    \end{equation}
    where $\lambda_i$ are the eigenvalues of $\mathcal{M}_{E,\Lambda}(\rho)$,

    \item[] \textit{Spectral norm:}
    \begin{equation}
    E_\Lambda(\rho) = \|\mathcal{M}_{E,\Lambda}(\rho)\|_2 := \max_{\|{\bf x}\|_2=1} \|\mathcal{M}_{E,\Lambda}(\rho){\bf x}\|_2,
    \label{SN}
    \end{equation}
    for ${\bf x}\in \mathbb{R}^{n}$, where $\|\cdot\|_2$ denotes the Euclidean norm.
\end{itemize}
The GSE admits network and geometric interpretations, which are discussed in the End Matter.

\textit{Property.---}
Given a faithful underlying bipartite entanglement measure $E$ and a subsystem family $\Lambda$, the corresponding multipartite GSE satisfies the following properties \cite{Horodecki2009, Vedral1997, Vidal2000, Donald2002}.

\begin{enumerate}
\item \textit{Nonnegativity:} 
$E_\Lambda(\rho)\geq 0$, and
$E_\Lambda(\rho)=0$ if and only if
$\mathcal{M}_{E,\Lambda}(\rho)=0$.

\item \textit{Consistency:} 
In the case of two parties, $\Lambda=\{I_1,I_2\}$, one obtains $E_\Lambda(\rho)=E_{I_1|I_2}(\rho)$, showing that the spectral entanglement coincides with the underlying bipartite entanglement measure.

\item \textit{Separability:}
The state $\rho$ is fully separable with respect to the $M$ associated subsystems specified by the index set $I_\Lambda$ in Eq.~\eqref{MDSS1} if and only if $E_\Lambda(\rho)=0$, provided that $\Lambda$ contains the subsystem blocks associated with every possible bipartition of $I_\Lambda$.

\item \textit{Permutation invariance:}
$E_\Lambda(\rho)$ is invariant under any permutation in $\Lambda$. This follows from the fact that a permutation of any pair of subsystems $I_k$ and $I_l$ in $\Lambda$ corresponds to swapping the $k$-th and $l$-th rows and the same columns of the entanglement matrix $\mathcal{M}_{E, \Lambda}(\rho)$, i.e.,
\begin{eqnarray}
\mathcal{M}_{E, P\Lambda}(\rho) = P \mathcal{M}_{E, \Lambda}(\rho) P^{T},
\end{eqnarray}
where $P$ is a permutation matrix. Such a transformation leaves the eigenvalues, and hence the GSE, unchanged. This property shows that the GSE is independent of the order of the subsystems within the subsystem family $\Lambda$.

\item \textit{Invariant under local unitaries:} Consider a local unitary transformation $\tilde{\rho} = \left[\left(\otimes_{i\in I_{\Lambda}}U_i\right)\otimes V\right]\rho\left[\left(\otimes_{i\in I_{\Lambda}}U_i\right)\otimes V\right]^{\dagger}$,
where each $U_i$ represents a unitary acting on the subsystem $\mathcal{H}_i$ for $i\in I_{\Lambda}$ and $V$ is a unitary on the rest of the system, i.e., $\otimes_{i\notin I_{\Lambda}}\mathcal{H}_i$. For such transformation, one can obtain
$\tilde{\rho}_{I_k I_l} = \left[\left(\otimes_{i\in I_k}U_i\right)\otimes\left(\otimes_{i\in I_l}U_i\right)\right]\rho_{I_k I_l}\left[\left(\otimes_{i\in I_k}U_i\right)\otimes\left(\otimes_{i\in I_l}U_i\right)\right]^{\dagger}$.$\ $Since the underlying bipartite entanglement measure is invariant under local unitary transformations \cite{Vedral1997, Horodecki2009, Donald2002}, so is the entanglement matrix and thus the GSE, i.e.,
\begin{equation}
E_\Lambda(\tilde{\rho}) = E_\Lambda(\rho).
\end{equation}

\item \textit{Convexity:} $E_\Lambda(\rho)$ satisfies the convexity relation
\begin{eqnarray}
E_\Lambda\left(\sum_{j} w_{j} \rho^{(j)}\right) \le \sum_{j} w_{j} E_\Lambda(\rho^{(j)}),
\label{CSE}
\end{eqnarray}
for any convex combination $\rho = \sum_{j} w_{j} \rho^{(j)}$ of a set of density matrices $\rho^{(j)}$ with weights $w_{j} \ge 0$ such that $\sum_{j} w_{j} = 1$. This follows from the linearity of the partial trace, the convexity of the underlying bipartite entanglement measure $E$, and the monotonicity and triangle inequality of the spectral norm \cite{Horn2012}. A detailed proof is provided in the End Matter.

\item \textit{Monotonicity:} 
$E_\Lambda(\rho)$ is monotone under LOCC, that is, for any quantum operation $\Phi$ given by local operations and classical communication, one has \cite{Vedral1997, Horodecki2009, Donald2002}
\begin{eqnarray}
E_\Lambda(\Phi[\rho]) \le E_\Lambda(\rho).
\label{MR}
\end{eqnarray}
The proof relies on the local-operation property of the partial trace, the monotonicity of the underlying bipartite entanglement measure $E$, and the monotonicity of the spectral norm with respect to entrywise ordering for nonnegative matrices \cite{Horn2012}. The detailed proof is provided in the End Matter.
\end{enumerate}

\textit{Generality of the construction}:
The graph-spectral construction is independent of the particular bipartite entanglement measure used to define the edge weights. Thus, different entanglement measures and subsystem families can be incorporated within the same framework to explore a range of multipartite settings.

Beyond the properties established above, GSE satisfies the following {\it inclusion monotonicity}, a fundamental feature underlying its multipartite structure. This property induces a natural hierarchy of correlations within a quantum system and provides a way to characterize the multipartite nature of GSE through the generalized residual entanglement introduced below.

\textbf{Lemma} (Inclusion monotonicity).
\textit{For families of nontrivial subsets $\Lambda_1 \subseteq \Lambda_2 \subseteq \mathcal{P}(I)$, the GSE satisfies
\begin{equation}
E_{\Lambda_1}(\rho) \le E_{\Lambda_2}(\rho),
\label{eq:lemma}
\end{equation}
for any quantum state $\rho$ on the Hilbert space in
Eq.~\eqref{eq:NPS}.}

{\it Proof.}
From the definition, the entanglement graph $G(\rho,\Lambda_1,E)$ is a subgraph of $G(\rho,\Lambda_2,E)$. Consequently, the corresponding entanglement matrix $\mathcal{M}_{E,\Lambda_1}(\rho)$ is a principal submatrix of $\mathcal{M}_{E,\Lambda_2}(\rho)$. The Eq.~\eqref{eq:lemma} then follows from   Eq.~\eqref{SRG} (equivalently, Eq.~\eqref{SN}) and the monotonicity of the spectral radius (equivalently, the spectral norm) for nonnegative matrices under principal submatrices~\cite{Horn2012}.

To demonstrate the multipartite character of the GSE measure, we examine its connection to the residual three-tangle~\cite{Coffman2000}. We show that, for a particular partitioning, the GSE is a monotonically increasing function of the residual three-tangle. This relation is analogous to the monotonic relationship between concurrence and entanglement of formation for two-qubit states~\cite{Wootters1998}.

Let us consider a three-qubit system with the total Hilbert space
\begin{eqnarray}
\mathcal{H}=\mathcal{H}_{A}\otimes\mathcal{H}_{B}\otimes\mathcal{H}_{C}
\label{trips}
\end{eqnarray}
For the particular choice of subsystem family $\Delta = \{\{A\}, \{B\}, \{C\}, \{B,C\}\}$,
a bipartite entanglement measure $E$, and a tripartite density matrix $\rho$, the corresponding entanglement matrix is
\begin{equation}
\mathcal{M}_{E,\Delta}(\rho) =
\begin{pmatrix}
0 & E_{A|B} & E_{A|C} & E_{A|BC} \\
E_{A|B} & 0 & E_{B|C} & 0 \\
E_{A|C} & E_{B|C} & 0 & 0 \\
E_{A|BC} & 0 & 0 & 0
\end{pmatrix}.
\label{eq:3qmatrix1}
\end{equation}

Following the three-qubit monogamy relation~\cite{Coffman2000, Osborne2006}, 
\begin{equation}
E_{A|B}^2 + E_{A|C}^2\le E_{A|BC}^2,
\label{eq:ckw0}
\end{equation}
the
three-tangle $\tau_{A|B|C}$, defined as
\begin{equation}
\tau_{A|B|C}=E_{A|BC}^2 - E_{A|B}^2 - E_{A|C}^2,
\label{eq:ckw1}
\end{equation}
quantifies tripartite entanglement for pure states. When $E$ is the concurrence \cite{Hill1997, Wootters1998, Rungta2001, Def-concurrence-negativity}, $\tau_{A|B|C}$ is known as the
residual tangle~\cite{Coffman2000}. A similar residual construction based on negativity \cite{Vidal2002, Def-concurrence-negativity} has been also introduced~\cite{Ou2007}. 

By substituting the definition in Eq.~\eqref{eq:ckw1} into the characteristic polynomial of the entanglement matrix in Eq.~\eqref{eq:3qmatrix1}, i.e., $P(\lambda)=\det(\lambda I-\mathcal{M}_{E,\Delta}(\rho))$, one obtains
\begin{eqnarray}
P(\lambda,\tau_{A|B|C})
&=& \lambda^4
- (2E_{A|B}^2 + 2E_{A|C}^2 + E_{B|C}^2 + \tau_{A|B|C}) \lambda^2\nonumber\\
&&- (2E_{A|B}E_{A|C}E_{B|C})\lambda\nonumber\\
&&+ E_{B|C}^2 (E_{A|B}^2 +E_{A|C}^2 + \tau_{A|B|C}).
\label{characteristic-polynomial}
\end{eqnarray}
All eigenvalues $\lambda_i$ of $\mathcal{M}_{E, \Delta}(\rho)$, including the largest eigenvalue
$\lambda_{\max}=\max\{\lambda_1, \lambda_2, \lambda_3, \lambda_4\}$
that defines the corresponding GSE as $E_\Delta(\rho)=\lambda_{\max}$, satisfy $P(\lambda_i,\tau_{A|B|C})=0$. Thus, the GSE is a function of the residual tangle $\tau_{A|B|C}$, implicitly defined by the relation
$P\left(E_\Delta(\rho),\tau_{A|B|C}\right)
=P(\lambda_{\max},\tau_{A|B|C})=0$.
Therefore, by applying the implicit function theorem, one obtains
\begin{equation}
\frac{\partial E_\Delta(\rho)}{\partial\tau_{A|B|C}}=\frac{\partial\lambda_{\max}}{\partial\tau_{A|B|C}}
= - \frac{\partial P/\partial \tau_{A|B|C}}{\partial P/\partial \lambda}
\bigg|_{\lambda=\lambda_{\max}}.
\label{ID}
\end{equation}

Taking the partial derivative of the characteristic polynomial in Eq.~\eqref{characteristic-polynomial} with respect to $\tau_{A|B|C}$ at $\lambda=\lambda_{\max}$, we obtain
\begin{equation}
\frac{\partial P}{\partial \tau_{A|B|C}}\Big|_{\lambda=\lambda_{\max}} =-\lambda_{max}^2 + E_{B|C}^2\le 0,
\label{EQN1}
\end{equation}
where the inequality follows from the Rayleigh quotient theorem \cite{Horn2012}, applied to the quadratic-form expression in Eq.~\eqref{eq:quadraticformmultipartiteE}. In particular, choosing the normalized vector ${\bf x}_0=\frac{1}{\sqrt{2}}(0,1,1,0)^T$, we obtain
\begin{eqnarray}
\lambda_{\max} = \max_{\|{\bf x}\|_2 = 1} {\bf x}^\top \mathcal{M}_{E, \Delta}(\rho) {\bf x}
\ge{\bf x}_0^\top \mathcal{M}_{E, \Delta}(\rho) {\bf x}_0
= E_{B|C},
\end{eqnarray}
showing that the partial derivative in Eq.~\eqref{EQN1} is nonpositive.

On the other hand, since $P(\lambda,\tau_{A|B|C})$ has a positive leading coefficient and all its roots are real eigenvalues of $\mathcal{M}_{E,\Delta}(\rho)$, it can be factorized as $P(\lambda,\tau_{A|B|C})=\prod_{i=1}^{4}(\lambda-\lambda_i)$. Hence, the derivative at the largest root satisfies
\begin{equation}
\frac{\partial P}{\partial \lambda}\Big|_{\lambda=\lambda_{\max}}
= \prod_{i\neq \max}(\lambda_{\max}-\lambda_i) > 0 .
\label{EQDN2}
\end{equation}

Combining Eqs.~\eqref{EQN1} and \eqref{EQDN2} with the implicit differentiation relation in Eq.~\eqref{ID}, we conclude
\begin{equation}
\frac{\partial E_\Delta(\rho)}{\partial\tau_{A|B|C}}
= \frac{\lambda_{\max}^2 - E_{B|C}^2}
{\prod_{i\neq \max}(\lambda_{\max}-\lambda_i)}
> 0.
\end{equation}
Therefore, the GSE $E_\Delta(\rho)$ is an increasing function of the three-tangle $\tau_{A|B|C}$. This demonstrates that the GSE is, in fact, a tripartite entanglement measure in this example integrating two- and three-tangles into a single unified measure. An example in the End Matter further demonstrates the multipartite character of the GSE beyond two-qubit marginal entanglement and residual three-tangle.

\begin{figure}[t]
    \centering \includegraphics[width=0.32\textwidth]{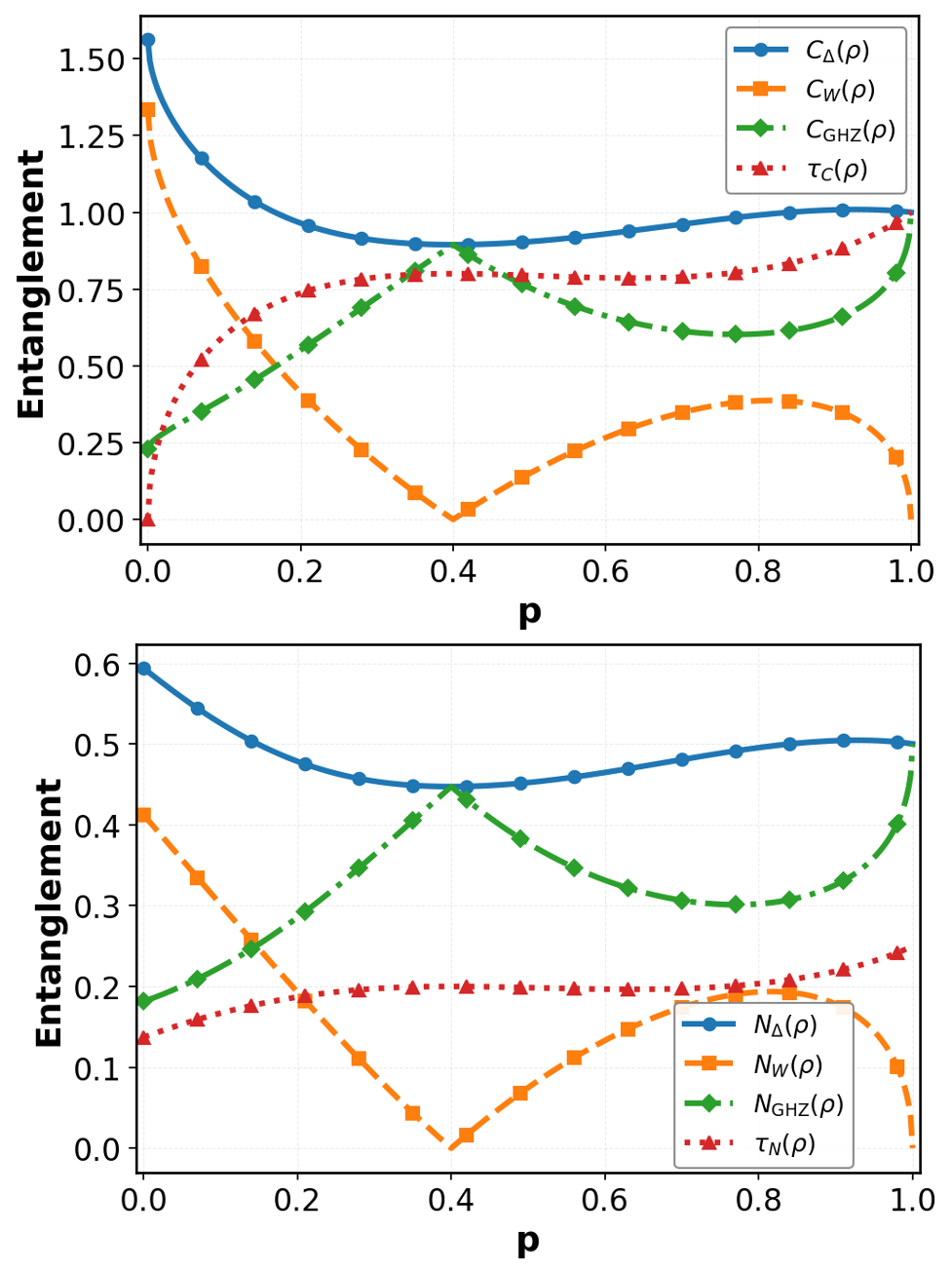}
\caption{The GSE $E_\Delta(\rho)$, its residual contributions $E_{\mathrm{W}}(\rho)$, $E_{\mathrm{GHZ}}(\rho)$, and the three-tangle $\tau_E(\rho)=\tau_{A|B|C}$ for the parametrized three-qubit state $\rho=\ket{\psi(p)}\bra{\psi(p)}$, where $\ket{\psi(p)}=\sqrt{p}\ket{\mathrm{GHZ}}+\sqrt{1-p}\ket{\mathrm{W}}$, shown as functions of the mixing parameter $p$. The upper and lower panels correspond to concurrence $C$ and negativity $N$, respectively, as the reference bipartite entanglement measure $E$. In both cases, the residual entanglement $\tau_E(\rho)=\tau_{A|B|C}$ shows qualitatively similar behavior to the spectral GHZ-type contribution $E_{\mathrm{GHZ}}(\rho)$. The GSE $E_\Delta(\rho)$ captures tripartite entanglement and unifies different contributions of residual entanglement.}
    \label{fig:SPEvsRE}
\end{figure}

Moreover, by inclusion monotonicity, we have
\begin{equation}
E_{\mathcal{S}_1}(\rho) \le E_\Delta(\rho),
\label{IP}
\end{equation}
where $\mathcal{S}_1 = \{\{A\}, \{B\}, \{C\}\}\subseteq \Delta$. Since the quantity $E_{\mathcal{S}_1}(\rho)$ captures only the bipartite entanglement between individual parties, Eq.\ \eqref{IP} provides a spectral form of the monogamy relation in Eq.~\eqref{eq:ckw0}. 
Based on the monogamy relation in Eq.~\eqref{eq:ckw0} and the residual-entanglement decomposition in Eq.~\eqref{eq:ckw1}, the GSE with respect to $\Delta$ can be decomposed into distinct contributions
\begin{equation}
E_{\mathrm{W}}(\rho) = E_{\mathcal{S}_1}(\rho), \quad
E_{\mathrm{GHZ}}(\rho) = E_\Delta(\rho) - E_{\mathcal{S}_1}(\rho).
\label{eq:W-GHZ-residual}
\end{equation}
Here, $\ket{\mathrm{W}} = \frac{1}{\sqrt{3}} \big(\ket{100} + \ket{010} + \ket{001}\big)$ and
$\ket{\mathrm{GHZ}} = \frac{1}{\sqrt{2}} \big(\ket{000} + \ket{111}\big)$
represent the two genuinely tripartite-entangled classes of three-qubit pure states, which are inequivalent under
stochastic LOCC \cite{Dur2000}. The three-tangle $E_{\mathrm{GHZ}}(\rho)$ provides a spectral analog of the residual tangle in Eq.~\eqref{eq:ckw1} and is therefore referred to as the \emph{residual graph-spectral three-tangle}.
Figure~\ref{fig:SPEvsRE} illustrates the relevant entanglement contributions for both concurrence~\cite{Wootters1998, Rungta2001, Def-concurrence-negativity} and negativity~\cite{Vidal2002, Def-concurrence-negativity}
. In both cases, the conventional residual tangle $\tau_{A|B|C}$ exhibits qualitatively similar behavior to the residual graph-spectral three-tangle $E_{\mathrm{GHZ}}(\rho)$. The GSE $E_\Delta(\rho)$ captures marginal bipartite, genuine tripartite, and residual contributions within a unified spectral characterization.

The monogamy relation and inclusion monotonicity express the same principle of entanglement distribution: extending a collection of partitions can only increase the associated entanglement. Thus, inclusion monotonicity implies a spectral analog of the monogamy inequality,
\begin{equation}
E_{\Lambda_1}(\rho) \le E_{\Lambda_2}(\rho),
\end{equation}
for any nontrivial partitions $\Lambda_1 \subseteq \Lambda_2 \subseteq \mathcal{P}(I)$. This naturally extends the notion of residual tangle and motivates the definition of the \textit{residual GSE},
\begin{equation}
\tau_{\Lambda_1, \Lambda_2}
= E_{\Lambda_2}(\rho)-E_{\Lambda_1}(\rho).
\label{eq:lemma22}
\end{equation}

To characterize multipartite entanglement across different scales, we introduce the nested hierarchy
\begin{equation}
\mathcal{S}_1 \subseteq \mathcal{S}_2 \subseteq \cdots \subseteq \mathcal{S}_{N-1} \subseteq \mathcal{P}(I),
\end{equation}
where $\mathcal{S}_n$ denotes the collection of all nonempty subsets of $I$ with cardinality at most $n$. The associated \emph{principal residual GSE} contributions are then defined by
\begin{equation}
\tau_n
= E_{\mathcal{S}_{n-1}}(\rho)-E_{\mathcal{S}_{n-2}}(\rho),
\qquad n=2,3,\ldots,N,
\label{eq:PREs}
\end{equation}
with the convention $E_{\mathcal{S}_0}(\rho)=0$.
Each $\tau_n$ quantifies the incremental contribution to the entanglement structure at the $n$-th level of the hierarchy in an $N$-body system. Eqs.~\eqref{eq:lemma22} and \eqref{eq:PREs} thus extend the notion of residual entanglement to general multipartite systems, providing a hierarchical characterization of their entanglement structure.

\begin{figure}[h]
    \centering \includegraphics[width=0.32\textwidth]{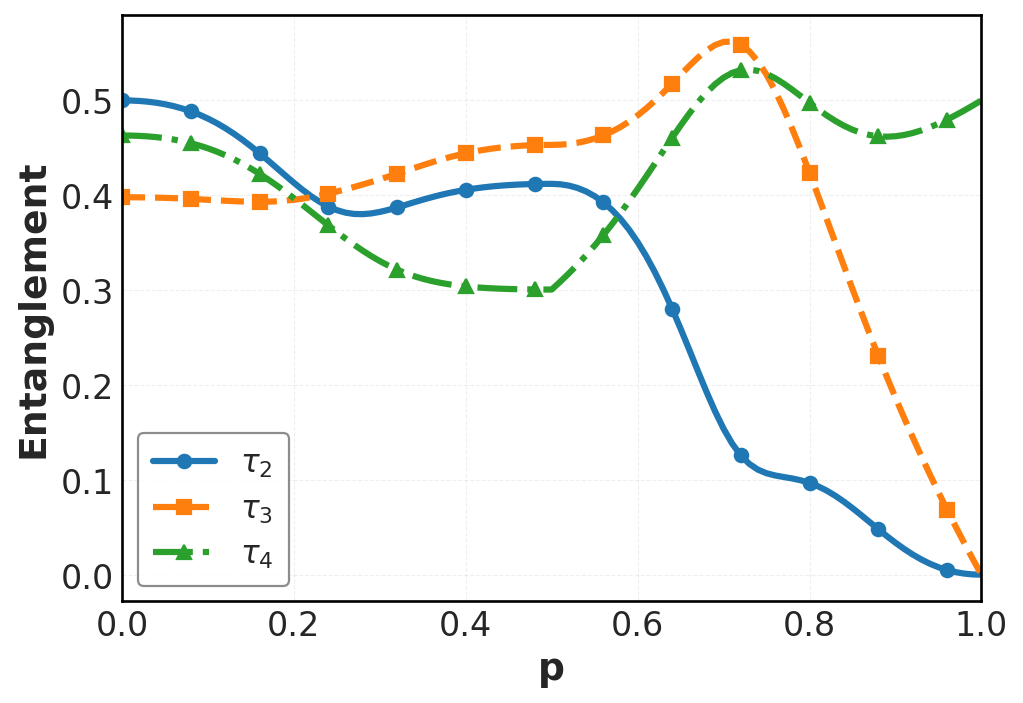}
\caption{Principal residual GSEs $\tau_2$, $\tau_3$, and $\tau_4$ of the four-qubit state $\rho=\ket{\phi(p)}\bra{\phi(p)}$, with $\ket{\phi(p)}$ given in Eq.~\eqref{phip}, as functions of the mixing parameter $p$. The changes in the ordering of $\tau_2$, $\tau_3$, and $\tau_4$ indicate transitions between distinct multipartite entanglement classes across the parameter range.}
\label{fig:PRE}
\end{figure}

Figure~\ref{fig:PRE} shows the principal residual GSE components $\tau_2$, $\tau_3$, and $\tau_4$ for the parametrized four-qubit state
$\rho=\ket{\phi(p)}\bra{\phi(p)}$,
where
\begin{equation}
\ket{\phi(p)}
=
\begin{cases}
\sqrt{1-2p}\ket{D_4^{(2)}}
+\sqrt{2p}\ket{W_3}\otimes\ket{0},
& 0\le p\le \tfrac12, \\[3mm]
\sqrt{2-2p}\ket{W_3}\otimes\ket{0}
+\sqrt{2p-1}\ket{\mathrm{GHZ}_4},
& \tfrac12 < p\le 1.
\end{cases}
\label{phip}
\end{equation}
This one-parameter family of states interpolates between the symmetric four-qubit Dicke state with two excitations, $\ket{D_4^{(2)}}
=\frac{1}{\sqrt{6}}\sum_{\pi}\ket{\pi(0011)}$ with sum running over all distinct permutations $\pi$ of the qubits, the state
$\ket{W_3}\otimes\ket{0}= 
\frac{1}{\sqrt{3}}
\left[
\ket{001}
+\ket{010}
+\ket{100}
\right]\otimes\ket{0}$, and the four-qubit GHZ state 
$\ket{\mathrm{GHZ}_4}=
\frac{1}{\sqrt{2}}
\left[
\ket{0000}
+\ket{1111}
\right]$. The crossings and reordering of the residual GSEs $\tau_2$, $\tau_3$, and $\tau_4$ in Fig.~\ref{fig:PRE} reveal a nontrivial redistribution of multipartite correlations as the parameter $p$ varies. The dominance of $\tau_2$ near $p=0$, $\tau_3$ near $p=1/2$, and $\tau_4$ near $p=1$ reflects distinct correlation hierarchies associated with the states $\ket{D_4^{(2)}}$, $\ket{W_3}\otimes\ket{0}$, and $\ket{\mathrm{GHZ}_4}$, respectively. These pronounced rearrangements of the residual GSEs provide signatures of distinct multipartite entanglement structures, consistent with the inequivalence of these states under stochastic LOCC~\cite{Verstraete2002}.

Existing multipartite entanglement measures employ different aggregation principles, including algebraic relations for residual constructions, minimization or maximization over bipartitions, geometric distances, and convex-roof constructions \cite{Coffman2000,MaGME2011,Wei2003,Eisert2001,Vidal2000,Ma2024}. In contrast, the present framework represents bipartite entanglement contributions as weights of an entanglement graph and aggregates them through the spectral radius of the associated entanglement matrix. This spectral aggregation reflects both the strength and collective connectivity of multipartite correlations, providing a distinct geometric and network-based characterization of multipartite entanglement. Moreover, the same graph-spectral construction accommodates different bipartite entanglement measures and subsystem-partition structures within a common framework.

We have introduced a unified graph-spectral framework for quantifying multipartite entanglement by representing quantum states as entanglement graphs for specified subsystem family  structures and analyzing the spectra of their corresponding entanglement matrices. The resulting graph-spectral entanglement measure satisfies the requirements of an entanglement measure and accommodates different bipartite entanglement measures and subsystem-partition structures within a common construction. It further provides a basis for a general multipartite monogamy structure, which in turn motivates the definition of residual graph-spectral entanglements for general multipartite states. Beyond quantification, the graph-spectral approach provides a distinct geometric and network perspective on multipartite quantum correlations, offering a new route for their analysis and classification in many-body quantum systems from both theoretical and experimental perspectives.



\section{Multipartite separability structure}
For a given multipartite quantum state $\rho$ on the Hilbert space in Eq.~\eqref{eq:NPS}, and a specified family $\Lambda=\{I_1,\ldots,I_n\}$ in Eq.~\eqref{MDSS}, $\rho$ is separable with respect to $\Lambda$ if its reduced density matrix on the subsystem $I_\Lambda:=\bigcup_{k=1}^{n}I_k=\{i_1,\ldots,i_M\}\subseteq I$ admits a convex decomposition into product states across any collection of disjoint subsystem blocks from $\Lambda$ that covers $I_\Lambda$. 
Explicitly, for every collection $I_{k_1},\ldots,I_{k_{n'}}\in\Lambda$ that is pairwise disjoint and satisfies $I_\Lambda=\bigcup_{l=1}^{n'}I_{k_l}$, the reduced density matrix can be written as
\begin{equation}
\rho_{I_{\Lambda}}=\sum_j p_j\rho_{I_{k_1}}^{(j)}\otimes\cdots\otimes\rho_{I_{k_{n'}}}^{(j)},
\end{equation}
where $p_j\geq 0$, $\sum_j p_j=1$, and $\rho_{I_{k_l}}^{(j)}$ are density matrices acting on the Hilbert space associated with subsystem block $I_{k_l}$.
In particular, when the family $\Lambda$ contains all individual parties, i.e., $\{\{i_1\},\ldots,\{i_M\}\}\subseteq\Lambda$, the above definition is equivalent to full $M$-partite separability, namely,
\begin{equation}
\rho_{I_\Lambda}
=\sum_j p_j
\rho_{i_1}^{(j)}\otimes\cdots\otimes
\rho_{i_M}^{(j)},
\end{equation}
where $\rho_{i_l}$ is the reduced density matrix on the subsystem Hilbert space $\mathcal{H}_{i_l}$ in Eq.~\eqref{eq:NPS} for each $l=1,\ldots,M$.
This definition of multipartite separability is the natural $M$-partite generalization of bipartite separability \cite{Horodecki2009}.

Although not necessary in general, the inclusion of all possible bipartitions in $\Lambda$ provides a sufficient condition for the equivalence between separability with respect to $\Lambda$ and the vanishing of GSE. In particular, if $\Lambda$ contains the subsystem blocks corresponding to all possible bipartitions of $I_\Lambda$, then $\rho$ is fully $M$-partite separable if and only if
$E_\Lambda(\rho)=0$.
This equivalence follows from the fact that separability across every bipartition is equivalent to full multipartite separability.
In this context, the graph-spectral construction provides a framework for characterizing multipartite entanglement by organizing the relevant bipartite entanglements into a common graph structure. Specifically, the subsystem blocks associated with the considered partitions are represented by graph vertices, while the bipartite entanglement across pairs of disjoint blocks defines the corresponding edge weights. The resulting entanglement matrix thus encodes both the strengths of the mutual quantum correlations and the connectivity of the subsystem blocks. Its spectral properties consequently provide a global characterization of the corresponding multipartite entanglement structure. In this sense, the multipartite character of the construction arises not merely from considering different bipartitions individually, but from their collective organization within a unified graph-spectral framework.
\vspace{-5mm}
\section{Network and Geometric interpretations }
\textit{Network interpretation.--}
The spectral radius of the entanglement graph captures its global connectivity, or "entanglement spread", across subsystem nodes and thus encodes the collective $M$-partite entanglement structure.

In the proposed entanglement graph (network), subsystem blocks represent vertices (nodes), while the bipartite entanglement between pairs of disjoint blocks defines the edges (links). The edge weights quantify the strengths of bipartite entanglement between nodes, while the spectral properties of the associated entanglement matrix characterize their collective organization and connectivity, thereby providing a global characterization of the multipartite entanglement structure.

Although both approaches employ graph structures, the entanglement network considered here differs conceptually from conventional physical quantum networks, as discussed, for instance, in Refs. \cite{Cirac1997, Briegel1998, Kimble2008, Simon2017, Epping2017, Wehner2018, Kraft2021}. In the latter, networks describe physical configurations of parties connected through quantum sources, often subject to source-independence constraints. In contrast, our network is constructed from the bipartite entanglement between subsystem nodes within a given multipartite quantum system. Its purpose is to capture how these entanglement correlations are interconnected and collectively contribute to the multipartite entanglement structure. Furthermore, the proposed GSE framework provides a general approach that is also useful for analysing partial and global entanglement structures in physical quantum networks, including quantum communication networks and quantum internet.
\\

\textit{Geometric interpretation}:
The norm representation in Eq.~\eqref{SN} provides a geometric interpretation of the GSE $E_\Lambda(\rho)$ as the largest stretching factor of the linear map
${\bf x}\mapsto \mathcal{M}_{E,\Lambda}(\rho){\bf x}$,
acting on the unit sphere $S^{n-1}$. Equivalently, it corresponds to the length of the major semiaxis of the ellipsoid $\mathcal{M}_{E,\Lambda}(\rho)[S^{n-1}]$, obtained as the image of the unit sphere under the action of the entanglement matrix. Here, $S^{n-1}\subset\mathbb{R}^n$ represents the space of normalized strengths of the $n$ entanglement links, rather than quantum state space. This geometric perspective  provides an intuitive way to visualize and analyze how different entanglement contributions, including bipartite and higher-order contributions, influence the overall geometry of the entanglement structure.
\vspace{-5mm}
\section{Proofs of the Properties}

\textit{Convexity:} For any convex combination $\rho=\sum_j w_j\rho^{(j)}$ of density matrices $\rho^{(j)}$, with $w_j\geq 0$ and $\sum_j w_j=1$, the GSE satisfies the convexity relation in Eq.~\eqref{CSE}, namely,
\begin{eqnarray}
E_\Lambda\left(\sum_j w_j\rho^{(j)}\right)
\leq \sum_j w_j E_\Lambda(\rho^{(j)}).
\label{eq:CSE1}
\end{eqnarray}
To prove this, first note that the linearity of the partial trace gives
\begin{eqnarray}
\rho_{I_k I_l}
=\sum_j w_j\rho^{(j)}_{I_k I_l},
\end{eqnarray}
for each pair of subsystems $I_k$ and $I_l$. Thus, by the convexity of the underlying bipartite entanglement measure $E$, we have
\begin{eqnarray}
E(\rho_{I_k I_l})
\leq \sum_j w_j E(\rho^{(j)}_{I_k I_l}).
\label{eq:inequality1}
\end{eqnarray}
Since $E(\rho_{I_k I_l})$ and $E(\rho^{(j)}_{I_k I_l})$ are the corresponding entries of the entanglement matrices $\mathcal{M}_{E,\Lambda}(\rho)$ and $\mathcal{M}_{E,\Lambda}(\rho^{(j)})$, respectively, and are nonnegative, the inequality in Eq.~\eqref{eq:inequality1} implies the following entrywise inequality
\begin{eqnarray}
0\leq \mathcal{M}_{E,\Lambda}(\rho)
\leq \sum_j w_j\mathcal{M}_{E,\Lambda}(\rho^{(j)}).
\end{eqnarray}
By the monotonicity of the spectral norm with respect to entrywise ordering for nonnegative matrices \cite{Horn2012}, we obtain
\begin{eqnarray}
\left\|\mathcal{M}_{E,\Lambda}(\rho)\right\|_2
\leq
\left\|\sum_j w_j\mathcal{M}_{E,\Lambda}(\rho^{(j)})\right\|_2.
\end{eqnarray}
Finally, using the definition in Eq.~\eqref{SN} and applying the triangle inequality for the spectral norm \cite{Horn2012}, we obtain
\begin{eqnarray}
\begin{aligned}
E_\Lambda(\rho)
&=\left\|\mathcal{M}_{E,\Lambda}(\rho)\right\|_2\\
&\leq \left\|\sum_j w_j\mathcal{M}_{E,\Lambda}(\rho^{(j)})\right\|_2\\
&\leq \sum_j w_j
\left\|\mathcal{M}_{E,\Lambda}(\rho^{(j)})\right\|_2=\sum_j w_j E_\Lambda(\rho^{(j)}),
\end{aligned}
\end{eqnarray}
which establishes the convexity relation in Eqs.~\eqref{eq:CSE1} and~\eqref{CSE}.

\textit{Monotonicity:} For any quantum operation $\Phi$ consisting of local operations and classical communication (LOCC), the follwoing inequalty is satisfed 
\begin{eqnarray}
E_\Lambda(\Phi[\rho])\leq E_\Lambda(\rho),
\label{eq:MR1}
\end{eqnarray}
as the LOCC monotonicity proery of GSE specified in Eq.~\eqref{MR}. 
To prove this, we first show that
\begin{eqnarray}
0\leq \mathcal{M}_{E,\Lambda}(\Phi[\rho])
\leq \mathcal{M}_{E,\Lambda}(\rho),
\label{MR1}
\end{eqnarray}
where the matrix inequalities are understood entrywise.

To this end, consider the induced map 
\begin{eqnarray}
\widetilde{\Phi}_{I_k I_l}[\rho_{I_k I_l}]
:=(\Phi[\rho])_{I_k I_l}=\operatorname{Tr}_{I\setminus(I_k\cup I_l)}(\Phi[\rho]).
\end{eqnarray}
on the corresponding reduced density matrices for any pair of subsystem blocks $I_k,I_l\in\Lambda$.
Since $\Phi$ is an LOCC operation across the individual parties and partial trace is a local quantum operation, the induced compost map is an LOCC operation across the bipartition $I_k|I_l$ \cite{Vidal2000}. Consequently, by the monotonicity of the underlying bipartite entanglement measure $E$ under LOCC, we obtain
\begin{eqnarray}
E\big((\Phi[\rho])_{I_k I_l}\big)
=E\big(\widetilde{\Phi}_{I_k I_l}[\rho_{I_k I_l}]\big)\leq&E(\rho_{I_k I_l}).
\end{eqnarray}
Since this inequality holds for every pair of subsystem blocks $I_k,I_l\in\Lambda$, and since two sides of the inequality represent entries of the corresponding entanglement matieces and are nonnegative, it follows that
\begin{eqnarray}
0\leq \mathcal{M}_{E,\Lambda}(\Phi[\rho])
\leq \mathcal{M}_{E,\Lambda}(\rho),
\end{eqnarray}
which proves Eq.~\eqref{MR1}.

Finally, using the definition of the GSE in Eq.~\eqref{SN} and the monotonicity of the spectral norm with respect to entrywise ordering for nonnegative matrices \cite{Horn2012}, we obtain
\begin{eqnarray}
E_\Lambda(\Phi[\rho])
=\left\|\mathcal{M}_{E,\Lambda}(\Phi[\rho])\right\|_2\leq\left\|\mathcal{M}_{E,\Lambda}(\rho)\right\|_2=E_\Lambda(\rho),
\end{eqnarray}
thereby establishing the LOCC monotonicity relation in Eqs.~\eqref{eq:MR1} and~\eqref{MR}.

\vspace{-5mm}
\section{Entanglement beyond pairwise concurrence and residual three-tangle}

Consider the following three-qubit mixed state, formed from a mixture of the three-qubit GHZ and W states introduced below Eq.~\eqref{eq:W-GHZ-residual}
\begin{equation}
\rho_{ABC}
=\frac{1}{2}\ket{\mathrm{GHZ}}\bra{\mathrm{GHZ}}
+\frac{1}{2}\ket{W}\bra{W}.
\label{eq:W-GHZ-state}
\end{equation}
The state is invariant under permutations of the three qubits, so its two-qubit reduced states are identical up to relabeling. In particular,
\begin{equation}
\rho_{AB}
=\frac{5}{12}\ket{00}\bra{00}
+\frac{1}{4}\ket{11}\bra{11}
+\frac{1}{3}\ket{\psi^+}\bra{\psi^+},
\end{equation}
where $\ket{\psi^+}=(\ket{01}+\ket{10})/\sqrt{2}$. Using the concurrence \cite{Wootters1998,Def-concurrence-negativity}, we obtain $C(\rho_{AB})=0$. By permutation symmetry, all pairwise concurrences therefore vanish
\begin{equation}
C(\rho_{AB})=C(\rho_{AC})=C(\rho_{BC})=0.
\end{equation}
Moreover, as shown in Ref.~\cite{Lohmayer2006}, the residual three-tangle of the mixed GHZ--W state in Eq.~\eqref{eq:W-GHZ-state}, defined through the convex-roof extension of the pure-state three-tangle in Eq.~\eqref{eq:ckw1}, also vanishes
\begin{equation}
\tau_{A|B|C}(\rho_{ABC})=0.
\end{equation}

Nevertheless, the state in Eq.~\eqref{eq:W-GHZ-state} is entangled. To characterize this entanglement within the graph-spectral framework, consider the subsystem family
\begin{equation}
\Lambda=\big\{
\{A\},\{B\},\{C\},
\{A,B\},\{A,C\},\{B,C\}
\big\},
\end{equation}
which includes all three bipartitions of $I_\Lambda=\{A,B,C\}$. Using the generalized concurrence $C$ \cite{Rungta2001,Def-concurrence-negativity} as the underlying bipartite entanglement measure, the corresponding entanglement matrix is
\begin{equation}
\mathcal M_{C,\Lambda}(\rho_{ABC})
=\frac{\sqrt{29}}{6}
\begin{pmatrix}
0&0&0&0&0&1\\
0&0&0&0&1&0\\
0&0&0&1&0&0\\
0&0&1&0&0&0\\
0&1&0&0&0&0\\
1&0&0&0&0&0
\end{pmatrix},
\label{eq:WGHZ-entanglement-matix}
\end{equation}
and consequently the graph-spectral entanglement is
\begin{equation}
C_\Lambda(\rho_{ABC})
=\frac{\sqrt{29}}{6}>0.
\label{eq:WGHZ-graph-spectral-entanglement}
\end{equation}
This confirms that the GHZ--W state exhibits entanglement beyond pairwise concurrences and the residual three-tangle.

The results in Eqs.~\eqref{eq:WGHZ-entanglement-matix} and~\eqref{eq:WGHZ-graph-spectral-entanglement} follow from the equal concurrences across the three bipartitions $A|BC$, $B|AC$, and $C|AB$. We explicitly calculate $C_{A|BC}$ below. Any pure-state decomposition of the state in Eq.~\eqref{eq:W-GHZ-state} has the form
\begin{equation}
\rho_{ABC}=\sum_jq_j\ket{\psi_j}\bra{\psi_j},
\qquad
q_j\geq0,\quad \sum_jq_j=1,
\label{eq:EM1}
\end{equation}
where each $\ket{\psi_j}$ lies in the support of $\rho_{ABC}$.  Every normalized pure state in this support can be expressed as
\begin{equation}
\ket{\psi}=\alpha\ket{\mathrm{GHZ}}+\beta\ket{W},
\qquad |\alpha|^2+|\beta|^2=1.
\end{equation}
Defining $x=|\alpha|^2$, the pure-state concurrence \cite{Def-concurrence-negativity} is
\begin{equation}
C_{A|BC}(\ket{\psi})
=\frac13\sqrt{8-4x+5x^2}.
\end{equation}
The function $f(x)=\frac13\sqrt{8-4x+5x^2}$ is convex, since
$f''(x)=\frac{12}{(8-4x+5x^2)^{3/2}}>0$.
Furthermore, the decompositions in Eq.~\eqref{eq:EM1} satisfy $\sum_j q_j|\alpha_j|^2=1/2$, where $\ket{\psi_j}=\alpha_j\ket{\mathrm{GHZ}}+\beta_j\ket{W}$ denotes each pure state in the decomposition.
It therefore follows from Jensen's inequality \cite{Rockafellar1970} that
\begin{equation}
C_{A|BC}(\rho_{ABC})
\geq f\left(\frac12\right)
=\frac{\sqrt{29}}{6}.
\end{equation}
This lower bound is attained by the decomposition
\begin{equation}
\rho_{ABC}
=\frac12\ket{\psi_+}\bra{\psi_+}
+\frac12\ket{\psi_-}\bra{\psi_-},
\quad
\ket{\psi_\pm}
=\frac{\ket{\mathrm{GHZ}}\pm\ket W}{\sqrt2},
\end{equation}
for which both pure states have concurrence $\sqrt{29}/6.$ Hence,
\begin{equation}
C_{A|BC}(\rho_{ABC})=\frac{\sqrt{29}}{6}.
\end{equation}
By permutation symmetry, we obtain
\begin{equation}
C_{A|BC}(\rho_{ABC})
=C_{B|AC}(\rho_{ABC})
=C_{C|AB}(\rho_{ABC})
=\frac{\sqrt{29}}{6}.
\end{equation}

This example points to a broader class of multipartite entangled mixed states, involving three or more parties, whose entanglement may not be fully captured by conventional measures such as pairwise concurrences and residual multipartite tangles. Investigating such states using GSE and comparing with existing results, such as those in Ref.~\cite{Lohmayer2006}, could help deepen our understanding of the complex entanglement structures of many-body mixed states and provide new ways to classify them.


\vspace{-5mm}

\begin{thebibliography}{99}
\bibitem{Einstein1935} A. B. Einstein, N. Podolsky, Rosen, Phys. Rev. {\bf 47}, 777 (1935).

\bibitem{Schrodinger1935} E. Schr\"odinger, Proc. Cambridge Philos. Soc. {\bf 31}, 555 (1935).

\bibitem{Schrodinger1936} E. Schr\"odinger, Proc. Cambridge Philos. Soc. {\bf 32}, 446 (1936).

\bibitem{Horodecki2009} R. Horodecki, P. Horodecki, M. Horodecki, K. Horodecki, Quantum entanglement, Rev. Mod. Phys. {\bf 81}, 865 (2009).

\bibitem{Nielsen2010} M. A. Nielsen, and I. L. Chuang, {\it Quantum computation and quantum information}, Cambridge university press (2010).

\bibitem{Giovannetti2011} V. Giovannetti, S. Lloyd, L. Maccone, Advances in quantum metrology, Nature photonics, {\bf 5}(4), 222-229 (2011).

\bibitem{Chitambar2019} E. Chitambar and G. Gour, Quantum resource theories, Rev. Mod. Phys. \textbf{91}, 025001 (2019).

\bibitem{Raussendorf2001}
R. Raussendorf and H. J. Briegel,
A One-Way Quantum Computer,
Phys. Rev. Lett. \textbf{86}, 5188 (2001).

\bibitem{Raussendorf2003}
R. Raussendorf, D. E. Browne, and H. J. Briegel,
Measurement-based quantum computation on cluster states,
Phys. Rev. A \textbf{68}, 022312 (2003).

\bibitem{Cirac1997} J. I. Cirac, P. Zoller, H. J. Kimble, and H. Mabuchi.
Quantum state transfer and entanglement distribution
among distant nodes in a quantum network. Phys. Rev.
Lett. {\bf 78}, 3221–3224 (1997).

\bibitem{Briegel1998} H.-J. Briegel, W. D\"ur, J. I. Cirac, and P. Zoller, Quantum Repeaters: The Role of Imperfect Local Operations in Quantum Communication, Phys. Rev. Lett. {\bf 81}, 5932 (1998).

\bibitem{Kimble2008} H. J. Kimble. The quantum internet. Nature {\bf 453}, 1023 (2008).

\bibitem{Simon2017} C. Simon. Towards a global quantum network. Nat. Phot.
11(11), 678 (2017).

\bibitem{Epping2017} M. Epping, H. Kampermann, C. Macchiavello, and D. Bru\ss{}, Multi-partite entanglement can speed up quantum key distribution in networks, New J. Phys. {\bf 19}, 093012 (2017).

\bibitem{Wehner2018} S. Wehner, D. Elkouss, and R. Hanson. Quantum internet: A vision for the road ahead. Science {\bf 362}(6412), eaam9288
(2018).

\bibitem{Navascues2020} M. Navascu\'es, E. Wolfe, D. Rosset, and A. Pozas-Kerstjens, Genuine network multipartite entanglement, Phys. Rev. Lett.  {\bf 125}(24), 240505 (2020).

\bibitem{Kraft2021} T. Kraft, S. Designolle, C. Ritz, N. Brunner, O. G\"uhne, and M. Huber, Quantum entanglement in the triangle network, Phys. Rev. A { \bf 103}, L060401 (2021).

\bibitem{Gaertner2007} S. Gaertner, C. Kurtsiefer, M. Bourennane, and H. Weinfurter, Experimental Demonstration of Four-Party Quantum Secret Sharing, Phys. Rev. Lett. {\bf 98}, 020503 (2007).

\bibitem{Zhou2018} Y. Zhou, J. Yu, Z. Yan, X. Jia, J. Zhang, C. Xie, and K. Peng, Quantum Secret Sharing Among Four Players Using Multipartite Bound Entanglement of an Optical Field, Phys. Rev. Lett. {\bf 121}, 150502 (2018).

\bibitem{Proietti2021} M. Proietti, J. Ho, F. Grasselli, P. Barrow, M. Malik, and A. Fedrizzi, Experimental quantum conference key agreement, Sci. Adv. {\bf 7} (23), eabe0395 (2021).

\bibitem{Yang2024} K.-X. Yang, Y.-L. Mao, H. Chen, X. Dong, J. Zhu, J. Wu, and Z.-D. Li, Experimental Measurement-Device-Independent Quantum Conference Key Agreement, Phys. Rev. Lett. {\bf 133}, 210803 (2024).

\bibitem{Ma2024} M. Ma, Y. Li, and J. Shang, Multipartite entanglement measures: A review, Fundam. Res. {\bf 5}, 2489-2499 (2025).

\bibitem{Dur2000} W. D\"ur, G. Vidal, and J. I. Cirac, Three qubits can be entangled in two inequivalent ways, Phys. Rev. A {\bf 62}, 062314 (2000).

\bibitem{Verstraete2002} F. Verstraete, J. Dehaene, B. De Moor, and H. Verschelde, Four qubits can be entangled in nine different ways,
Phys. Rev. A {\bf 65}, 052112 (2002).

\bibitem{Coffman2000} V. Coffman, J. Kundu, W. K. Wootters, Distributed entanglement, Phys. Rev. A {\bf 61}, 052306 (2000).

\bibitem{Osborne2006} T. J. Osborne, and F. Verstraete,  General Monogamy Inequality for Bipartite Qubit Entanglement, Phys. Rev. Lett. {\bf 96}, 220503 (2006).

\bibitem{Lohmayer2006} R. Lohmayer, A. Osterloh, J. Siewert, and A. Uhlmann, Entangled Three-Qubit States without Concurrence and Three-Tangle, Phys. Rev. Lett. {\bf 97}, 260502 (2006).

\bibitem{MaGME2011} Z.-H. Ma, Z.-H. Chen, J.-L. Chen, C. Spengler, A. Gabriel, and M. Huber, Measure of genuine multipartite entanglement with computable lower bounds, Phys. Rev. A \textbf{83}, 062325 (2011).

\bibitem{Eisert2001} J. Eisert and H. J. Briegel, Schmidt measure as a tool for quantifying multiparticle entanglement, Phys. Rev. A \textbf{64}, 022306 (2001).

\bibitem{Wei2003} T.-C. Wei and P. M. Goldbart, Geometric measure of entanglement and applications to bipartite and multipartite quantum states, Phys. Rev. A \textbf{68}, 042307 (2003).

\bibitem{Meyer2001} D. A. Meyer and N. R. Wallach, Global entanglement in multiparticle systems, J. Math. Phys. \textbf{43}, 4273 (2002).

\bibitem{Guhne2009} O. G\"uhne and G. T\'oth, Entanglement detection, Phys. Rep. {\bf 474}, 1-75 (2009).

\bibitem{Jungnitsch2011} B. Jungnitsch, T. Moroder, and O. Gühne, Taming multiparticle entanglement, Phys. Rev. Lett. \textbf{106}, 190502 (2011).

\bibitem{Wootters1998} W. K. Wootters, Entanglement of formation of an arbitrary state of two qubits, Phys. Rev. Lett. {\bf 80}, 2245-2248 (1998).

\bibitem{Hill1997} S. A. Hill and W. K. Wootters, Entanglement of a pair of quantum bits, Phys. Rev. Lett. {\bf 78}, 5022 (1997).

\bibitem{Rungta2001} P. Rungta, V. Bu{\v{z}}ek, C. M. Caves, M. Hillery, and G. J. Milburn, Universal State Inversion and Concurrence in Arbitrary Dimensions, Phys. Rev. A \textbf{64}, 042315 (2001).

\bibitem{Vedral1998} V. Vedral and M. B. Plenio, Entanglement measures and purification procedures, Phys. Rev. A {\bf 57}, 1619 (1998).


\bibitem{Vidal2002} G. Vidal, and R. F. Werner, Computable measure of entanglement, Phys. Rev. A {\bf 65}, 032314 (2002).

\bibitem{Christandl2004} M. Christandl, A. Winter, “Squashed entanglement”: An additive entanglement measure, J. Math. Phys. {\bf 45}(3), 829-840 (2004).

\bibitem{Plenio2007} M. B. Plenio and V. Sh. Virmani, “An introduction to entanglement measures,” Quantum Inf. Comput. {\bf 7}, 1–51 (2007).

\bibitem{Vedral1997} V. Vedral, M.B. Plenio, M.A. Rippin, P. L. Knight, Quantifying Entanglement, Phys. Rev. Lett. {\bf 78}, 2275 (1997).

\bibitem{Vidal2000} G. Vidal, Entanglement monotones, J. Mod. Opt. {\bf 47}, 355-376 (2000).

\bibitem{Donald2002} M. J. Donald, M. Horodecki, and O. Rudolph, The uniqueness theorem for entanglement measures, J. Math.
Phys. {\bf 43}, 4252 (2002).

\bibitem{Gour2013} G. Gour and N. R. Wallach, Classification of multipartite entanglement of all finite dimensionality,
Phys. Rev. Lett. {\bf 111}, 060502 (2013).

\bibitem{Toth2012} G. Tóth, Multipartite entanglement and high-precision metrology, Phys. Rev. A \textbf{85}, 022322 (2012).

\bibitem{Friis2018} N. Friis, O. Marty, C. Maier, C. Hempel, M. Holzäpfel, P. Jurcevic, M. B. Plenio, M. Huber, C. Roos, R. Blatt, and B. Lanyon, Observation of entangled states of a fully controlled 20-qubit system, Phys. Rev. X \textbf{8}, 021012 (2018).

\bibitem{Pu2018} Y. Pu, Y. Wu, N. Jiang, W. Chang, H. Zhang, S. Zhang, and L. Duan, Experimental entanglement of 25 individually accessible atomic quantum interfaces, Sci. Adv. \textbf{4}, eaar3931 (2018).

\bibitem{Zarkeshian2017} P. Zarkeshian, C. Deshmukh, N. Sinclair, S. K. Goyal, G. H. Aguilar, P. Lefebvre, M. Grimau Puigibert, V. B. Verma, F. Marsili, M. D. Shaw, S. W. Nam, K. Heshami, D. Oblak, W. Tittel, and C. Simon, Entanglement between more than two hundred macroscopic atomic ensembles in a solid, Nat. Commun. {\bf 8}, 906 (2017).

\bibitem{Frowis2017} F. Fr\"owis, P. C. Strassmann, A. Tiranov, C. Gut, J. Lavoie, N. Brunner, F. Bussi\`eres, M. Afzelius, and N. Gisin, Experimental certification of millions of genuinely entangled atoms in a solid, Nat. Commun. {\bf 8}, 907 (2017).

\bibitem{Lu2018} H. Lu, Q. Zhao, Z.-D. Li, X.-F. Yin, X. Yuan, J.-C. Hung, L.-K. Chen, L. Li, N.-L. Liu, C.-Z. Peng, Y.-C. Liang, X. Ma, Y.-A. Chen, and J.-W. Pan, Entanglement Structure: Entanglement Partitioning in Multipartite Systems and Its Experimental Detection Using Optimizable Witnesses, Phys. Rev. X \textbf{8}, 021072 (2018).

\bibitem{Horn2012} R. A. Horn and Ch. R. Johnson, \emph{Matrix Analysis}, Cambridge University Press, 2nd Edition (2012).

\bibitem{note} One may normalize the spectral entanglement in Eq.~\eqref{eq:multipartiteE} to unity using the Gers\v{s}gorin disc theorem \cite{Horn2012}, and define $E_\Lambda(\rho) = \frac{E(\rho)}{\max_{k}\sum_{l} m_{kl}}$. However, in this work we omit this normalization factor for simplicity.

\bibitem{Def-concurrence-negativity}
For a joint pure state $\rho_{12}=\ket{\psi}\bra{\psi}$ of a $d_1\times d_2$ quantum system, the generalized concurrence (I-concurrence) across the bipartition $1|2$ is defined as $C(\psi)=\sqrt{2\left(1-\operatorname{Tr}(\rho_1^2)\right)}=2\sqrt{\det\rho_1}$, where $\rho_1=\operatorname{Tr}_2(|\psi\rangle\langle\psi|)$ is the reduced density matrix of the first subsystem. For mixed states, the I-concurrence is defined through the convex-roof extension, namely, as the minimum average I-concurrence over all pure-state ensemble decompositions of the joint density operator, i.e., $C(\rho_{12})=\min_{\{p_i,\ket{\psi_i}\}}\sum_i p_i C(\ket{\psi_i})$, where $\rho_{12}=\sum_i p_i\ket{\psi_i}\bra{\psi_i}$~\cite{Rungta2001}. For two-qubit systems, the I-concurrence coincides with concurrence, for which an explicit analytical expression is available for mixed states~\cite{Wootters1998}.
The negativity across the bipartition $1|2$ for a joint density operator $\rho_{12}$ is defined as $N(\rho_{12})=\frac{\|\rho_{12}^{T_1}\|_1-1}{2}$, where $\rho_{12}^{T_1}$ denotes the partial transpose with respect to the first subsystem, and $\|X\|_1=\operatorname{Tr}\sqrt{X^\dagger X}$ is the trace norm~\cite{Vidal2002}. For three-qubit states, these definitions can be applied to the bipartitions $A|BC$, $B|AC$, and $C|AB$.

\bibitem{Ou2007}
Y.-C. Ou and H. Fan,
Monogamy inequality in terms of negativity for three-qubit states,
Phys. Rev. A \textbf{75}, 062308 (2007).

\bibitem{Rockafellar1970} R. T. Rockafellar, \emph{Convex Analysis}, Princeton University Press (1970).


\vspace{5pt}
\onecolumngrid

\begin{center}
\vspace{10pt}
\textbf{End Matter}
\end{center}
\vspace{10pt}
\twocolumngrid
\end{thebibliography}
\end{document}